\documentclass[english,pre,showpacs,floatfix]{revtex4}
\usepackage[T1]{fontenc}
\usepackage[latin1]{inputenc}
\usepackage{babel}
\usepackage{epsfig}
\usepackage{amsmath}
\usepackage{amssymb}

\DeclareMathOperator{\cosech}{cosech}
\DeclareMathOperator{\sech}{sech}
\DeclareMathOperator{\dn}{dn}
\DeclareMathOperator{\sn}{sn}
\DeclareMathOperator{\E}{E}
\DeclareMathOperator{\K}{K}

\begin{document}

\title
{\bf From collective periodic running states to completely chaotic synchronised states in coupled particle dynamics}
\author{ D. Hennig, A.D. Burbanks, A.H. Osbaldestin, and C. Mulhern}
\medskip
\medskip
\medskip
\affiliation{Department of Mathematics, University of
Portsmouth,\\Portsmouth, PO1 3HF, UK}

\begin{abstract}
\noindent We consider the damped and driven dynamics of two interacting particles evolving in a symmetric and spatially periodic potential. The latter is exerted to a time-periodic modulation of its inclination. Our interest is twofold: Firstly we deal with the issue of chaotic motion in the higher-dimensional phase space. To this end a homoclinic Melnikov analysis is utilised assuring the presence of transverse homoclinic orbits and homoclinic bifurcations for weak coupling allowing also for the emergence of hyperchaos.  In contrast, we also prove that the time evolution of the two coupled particles attains a 
completely synchronised (chaotic) state for strong enough coupling  between them. The resulting `freezing of dimensionality' rules out the occurrence of  hyperchaos. Secondly we address  coherent collective particle transport provided by regular periodic motion. A subharmonic Melnikov analysis is utilised to investigate persistence of periodic orbits. For directed particle transport mediated by rotating  periodic motion we present exact results regarding the collective character of the running solutions entailing the emergence of a current. We  show that coordinated energy exchange between the particles takes place in such a manner that they are enabled to 
overcome ---\,one particle followed by the other\,--- consecutive barriers of the periodic  potential resulting in collective directed motion.
\end{abstract}


\maketitle

\section{Lead Paragraph}

{\bf{ The study of transport phenomena has attracted considerable
interest over the years due to its relevance in many physical
situations, which are often modelled on the basis of
one-dimensional particle motion in a spatially periodic
potential. If the particles, in addition to their motion in the  periodic potential,
interact, cooperative effects not found in
situations of individual particle motion, may arise. 
The objective of the current work is to investigate the conditions
under which it is possible to generate a directed flow along with
collective motion in a system of two coupled particles. To be precise, we
study the transport of two particles interacting via a harmonic bond force.
The individual dynamics of uncoupled particle motion, evolving in a symmetric and spatiallay periodic potential whose inclination is time-periodically modulated, is chaotic, preventing the emergence of directed particle motion. 
For the dynamics of the coupled particles we elucidate the possible scenario in
which the energy exchange between the particles proceeds in such a
well-coordinated manner that the particles move separately from one
well into the next, one following the other, resulting in directed 
motion.
Notably, we demonstrate, that the dynamics of the coupled particles exhibits coherent collective directed motion provided by transporting attractors. Furthermore, it is shown that for suitable coupling strengths between the particles, a completely synchronised chaotic state is attained confining the dynamics in phase space as a result of `freezing of dimensionality'.}}

\section{Introduction}

Transport phenomena play a fundamental role in many physical systems. For a number of applications, including Josephson junctions \cite{Josephson}, charge density waves \cite{charge}, superionic conductors \cite{superionic}, rotation of dipoles in external fields \cite{dipoles}, phase-locked loops \cite{loops} and diffusion of dimers on surfaces \cite{surfaces}-\cite{Fusco} to quote a few, transport is based on the dynamics evolving in  spatially periodic potential landscapes.
When an additional external time-periodic modulation is applied to the periodic  potential, 
interesting effects such as phase-locking, hysteresis \cite{hysteresis} and stochastic resonance \cite{stochastic} are found. Recent investigations have dealt with the Hamiltonian dynamics of individual particles evolving in a periodic
potential whose inclination is time-periodically varied by
 a weak external monochromatic modulation field \cite{Yevtushenko,Soskin,single}. Remarkably, for the corresponding one-and-a-half degree of freedom Hamiltonian system it has been
demonstrated that adiabatic modulations of the slope of the potential lead to the generation of transient transport dynamics related with enormous directed particle flow. An explanation for this behaviour has been given in terms of the underlying phase space structure
of the externally driven one degree of freedom system 
promoting the motion in ballistic channels \cite{ballistic}. 
Recently interest in the collective transport dynamics of interacting particles in periodic potential landscapes has grown \cite{PhysicaD}-\cite{JPhys}.

In the present paper we consider the motion of two damped, time-periodically driven and coupled particles in a periodic potential. Our aim is to demonstrate that depending on the coupling strength between the two particle subsystems a rich dynamics results ranging from 
hyperchaos, stable periodic running states mediating coherent particle transport, to a completely synchronised chaotic state.
Motion takes place in a five-dimensional phase space for which details of its intricate structures remain elusive not least due to the higher dimensionality.
Whether in systems with a larger number of (microscopic) degrees of freedom such (macroscopic) behaviour as collective motion leading to a directed flow emanates from higher dimensional dynamics is not obvious as also more complex forms of unstable motion, namely hyperchaos may emerge \cite{hyperchaos}. We investigate the case when (already)  the dynamics of a single (uncoupled) particle is characterised by deterministic chaos. In this context several questions arise: namely (i) if the   chaotic motion persists in the coupled dynamics or, (ii) may the coupling drive the system's dynamics into a state of synchronisation, and (iii)  may chaos even become suppressed (see \cite{Pecora},\cite{Arkadi})? 

On the other hand, regarding coherent collective transport leading to the emergence of a directed flow it is paramount that a regime of regular periodic motion exists which is  characterised by transporting attractors. Such a regime of regular periodic motion needs to emerge from the coupled dynamics (we recall that a single, uncoupled system exhibits chaos).
These are the problems which we address. The paper is organised as follows: In the next section we introduce the model of the coupled particle system. Section \ref{section:Melnikov} is devoted to a  Melnikov analysis
related to homoclinic motion in the system. In the subsequent section we apply  a subharmonic Melnikov method to prove the persistence of periodic orbits in the coupled dynamics. In section \ref{section:features} we present exact results regarding the collective particle transport mediated by periodic running solutions. The ensuing current is considered in section 
\ref{section:current}. In \ref{section:complete} we discuss the synchronisation of the coupled particle system. We prove that for overcritical coupling strength a completely synchronised state is attained. Finally we summarise our results and give a brief outlook.

\section{The coupled particle system}\label{section:model}
We study the dynamics of two damped, coupled particles evolving in a spatially symmetric  (washboard) potential of spatial period $L=1$, with equations of motion given by
\begin{eqnarray}
\ddot{q}_1&=&-\sin(2\pi q_1)-\gamma \dot{q}_1-\,F\sin(\Omega\, t +\theta_0)-\kappa(q_1-q_2)\,\,,\label{eq:q1}\\
\ddot{q}_2&=&-\sin(2\pi q_2)-\gamma \dot{q}_2-F\sin(\Omega\, t +\theta_0)+\kappa (q_1-q_2)\,\,.\label{eq:q2}
\end{eqnarray}
An external time-dependent modulation of amplitude $F$, frequency $\Omega$ and phase $\theta_0$  serves for periodic modulations of the inclination of the washboard potential.

We emphasise that the  system is unbiased in the sense that the
forces averaged over time and space vanish, i.e.
\begin{equation}
\int\limits_{0}^{1}dq_i\int\limits_0^{T=2\pi/\Omega}dt
\frac{\partial V(q_1,q_2,t)}{\partial
q_i}=0\,,\,\,\,i=1,2.
\end{equation}
where the potential is given by
\begin{equation}
V(q_1,q_2,t)=\frac{1}{2\pi}(2-\cos(2\pi q_1)-\cos(2\pi q_2))+F\sin({\Omega \,t+\theta_0})(q_1+q_2)\,.
\end{equation}
Further, notice the exchange symmetry of the system $(\dot{q}_1,q_1) \leftrightarrow (\dot{q}_2,q_2)$.

\begin{figure}
\includegraphics[scale=1.0]{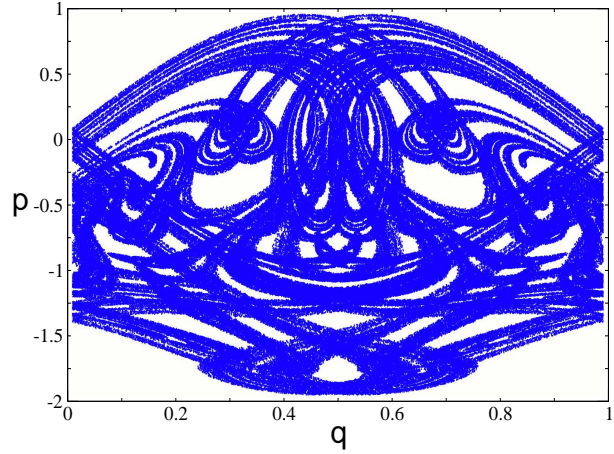}
\caption{Strange attractor for a single particle system, i.e. $\kappa=0$. The remaining parameter values are given by $\Omega=2.25$, $F=1.3$, $\theta_0=0$, and $\gamma=0.1$.} \label{fig:attractor}
\end{figure}

Before we embark on our study of the coupled particle dynamics we
briefly consider the dynamics of a single damped and driven particle system, i.e. $\kappa=0$.  The parameter values, $\Omega=2.25$, $F=1.3$, and $\theta_0=0$,  are chosen such that the dynamics is chaotic.  To illustrate the phase flow
we utilise  a Poincar\'{e} map using the period of
the external force, $T_0=2\pi/\Omega$, as the
stroboscopic time. The system of equations of motion was
integrated numerically and omitting a transient phase, points were
set in the map at times being multiples of the period duration
$T_0$. The resulting strange attractor is shown in the $(p=\dot{q},q)-$plane in Fig.~\ref{fig:attractor}.

\section{Melnikov-analysis}\label{section:Melnikov}

In this section we focus our interest on  the influence of the coupling between the particles on the character of the dynamics. In particular we are interested in a more complex form of unstable motion which may arise in the five-dimensional phase space of the system (\ref{eq:q1}),(\ref{eq:q2}), namely hyperchaos. 

We perform a Melnikov-analysis related to homoclinic motions in system (\ref{eq:q1}),(\ref{eq:q2}) for weak coupling, weak forcing, and weak damping which is expressed as
\begin{eqnarray}
\ddot{q}_1+\sin(2\pi q_1)&=&\epsilon\left(-\gamma \dot{q}_1-\,F\sin(\Omega\, t +\theta_0)-\kappa(q_1-q_2)\right)\,\,,\label{eq:q1eps}\\
\ddot{q}_2+\sin(2\pi q_2)&=&\epsilon \left(-\gamma \dot{q}_2-F\sin(\Omega\, t +\theta_0)+\kappa (q_1-q_2)\right)\,\,,\label{eq:q2eps}
\end{eqnarray}
where the parameter $\epsilon \ll 1$ indicates the perturbative character of the expressions on the r.h.s. in Eqs.\,(\ref{eq:q1eps}),(\ref{eq:q2eps}).
For $\epsilon=0$, two unperturbed pendula are obtained 
\begin{equation}
\dot{p}_i=-\sin(2\pi q_i)\,,\qquad \dot{q}_i=p_i\,,\qquad i=1,2\,.
\end{equation}
There exist hyperbolic fixed points given by
\begin{equation}
(\bar{p}_i,\bar{q}_i)=\left(0,\pm \frac{1}{2}\right)\,,\qquad i=1,2\,.
\end{equation}
Thus viewed in the full $p_1-q_1-p_2-q_2-\theta$ space the system has a hyperbolic periodic orbit
\begin{equation}
{\mathcal{M}}=(\bar{p}_1,\bar{q}_1,\bar{p}_2,\bar{q}_2,\theta(t))
=(0,\pm\frac{1}{2},0,\pm\frac{1}{2},\Omega t+\theta_0)\,.
\end{equation}
This hyperbolic periodic orbit is connected to itself by two pairs of homoclinic trajectories
given by
\begin{eqnarray}
& &\left(p^{\pm}_{1h}(t),q^{\pm}_{1h}(t),p^{\pm}_{2h}(t),q^{\pm}_{2h}(t),\theta(t)\right)=\nonumber\\
&=&\left(\pm\sqrt{\frac{2} {\pi}}\sech\left( \sqrt{2 \pi} t\right), \pm \frac{1}{\pi} \sin^{-1}\left[\tanh\left(\sqrt{2\pi} t\right)\right],\pm\sqrt{\frac{2}{\pi}}\sech\left( \sqrt{2 \pi} t\right),\right.\nonumber\\
 & & \left.\pm \frac{1}{\pi}\sin^{-1}\left[\tanh\left(\sqrt{2\pi} t\right)\right],\Omega t+\theta_0 \right)\label{eq:homoclinic}\,.
\end{eqnarray}
${\mathcal{M}}$ has three-dimensional stable and unstable manifolds which coincide along four  three-dimensional sets of homoclinic orbits, denoted by $\Gamma^{\pm}$, which can be parametrised as
\begin{eqnarray}
\Gamma^{\pm}&=&\left\{\left(p^{\pm}_{1h}(-\tau_1),q^{\pm}_{1h}(-\tau_1),p^{\pm}_{2h}(-\tau_2),q^{\pm}_{2h}(-\tau_2),\theta_0\right) \right.\nonumber\\& & \left. \in \mathbb{R}^1\times \mathbb{T}^1 \times \mathbb{R}^1\times \mathbb{T}^1 \times \mathbb{T}^1  \mid (\tau_1,\tau_2,\theta_0)\in
\mathbb{R}^1\times \mathbb{R}^1 \times \mathbb{T}^1\right\}\,.
\end{eqnarray}
The real parameters $\tau_{1,2}$ determine the position on the homoclinic orbits.
To determine whether the stable and unstable manifolds intersect transversely we compute the Melnikov integrals (for details concerning the Melnikov method in higher dimensions see  \cite{Wiggins}-\cite{Yagasaki1})
\begin{eqnarray}
M^{\pm}_i(\tau_1,\tau_2)&=&\int_{-\infty}^{\infty}\left\{p^{\pm}_{ih}(t-\tau_i)\left[-\gamma
p^{\pm}_{ih}(t-\tau_i)-F\sin(\Omega t+\theta_0)\right.\right.\nonumber\\
&+&\left.\left.\kappa \left( q^{\pm}_{i+1h}(t-\tau_{i+1})-q^{\pm}_{ih}(t-\tau_{i})\right)\right]\right\}dt\,.
\end{eqnarray}
The integration is performed over the homoclinic trajectories given in Eqs.\,(\ref{eq:homoclinic}) yielding the following integrals
\begin{eqnarray}
&=&\int_{-\infty}^{\infty}\left\{ \pm\sqrt{\frac{2}{\pi}}\sech\left( \sqrt{2 \pi} (t-\tau_i)\right)\left[\mp \gamma \sqrt{\frac{2}{\pi}}\sech\left( \sqrt{2\pi} (t-\tau_i)\right)\right.\right.\nonumber\\
&-&F\sin(\Omega t+\theta_0)+\kappa \left(
\pm \frac{1}{\pi}\sin^{-1}\left[\tanh\left(\sqrt{2 \pi} (t-\tau_{i+1})\right)\right]\right.\nonumber\\
&\mp&\left.\left.\left. \frac{1}{\pi}\sin^{-1}\left[\tanh\left(\sqrt{2\pi} (t-\tau_i)\right)\right]\right)\right]\right\}dt\,,\,\,\, i=1,2\,,\,\,\,\tau_{3}=\tau_1 ,\,\,\,{\rm and} \,\,\,\, q_{3 h}^{\pm}=q^{\pm}_{1 h}\,.
\end{eqnarray}

Using the method of residues we obtain for the case of equal signs
\begin{equation}
M^{(\pm,\pm)}_1(\tau_1,\tau_2)=-2\gamma  \sqrt{\frac{2}{\pi^3}} \mp F\,\sech\left(\sqrt{\frac{\pi}{2}}\,\frac{\Omega}{2} \right) \sin(\Omega \tau_1)+\frac{\kappa}{\pi^2}I(\Delta \tau)
\,,\label{eq:Mpp1}
\end{equation}
and
\begin{equation}
M^{(\pm,\pm)}_2(\tau_1,\tau_2)=-2\gamma \sqrt{\frac{2}{\pi^3}}\mp F\,\sech\left(\sqrt{\frac{\pi}{2}}\,\frac{\Omega}{2} \right) \sin(\Omega \tau_2)-\frac{\kappa}{\pi^2}I(\Delta \tau)\,.\label{eq:Mpp2}
\end{equation}

For unequal signs we obtain
\begin{equation}
M^{(\pm,\mp)}_1(\tau_1,\tau_2)=-2\gamma  \sqrt{\frac{2}{\pi^3}}\mp F\,\sech\left(\sqrt{\frac{\pi}{2}}\,\frac{\Omega}{2} \right) \sin(\Omega \tau_1)-\frac{\kappa}{\pi^2}I(\Delta \tau)\,,\label{eq:Mpm1}
\end{equation}
and
\begin{equation}
M^{(\pm,\mp)}_2(\tau_1,\tau_2)=-2\gamma  \sqrt{\frac{2}{\pi^3}} \pm F\,\sech\left(\sqrt{\frac{\pi}{2}}\,\frac{\Omega}{2} \right) \sin(\Omega \tau_2)+\frac{\kappa}{\pi^2}I(\Delta \tau)\,.\label{eq:Mpm2}
\end{equation}

The function $I( \Delta \tau)$ is determined by the integral
\begin{equation}
I(\Delta \tau)=\int_{-\infty}^{\infty} \frac{\sin^{-1}\left[\tanh(t+ \Delta \tau)\right]}{\cosh(t)}dt\,,\label{eq:integral}
\end{equation}
where
\begin{equation}
 \Delta \tau= \sqrt{2\pi}(\tau_1 -\tau_2)\,.
\end{equation}

Denoting $t=\Delta \tau$, the function $I(t)$ has the following properties
\begin{equation}
I(0)=0\,,\,\,\,
I(t)=-I(-t)\,,\,\,\,
\frac{dI(t)}{dt}>0\,,
\,\,\,{\rm and}\,\,\,\max_{t \in {\mathbb{R}}}\mid I (t)\mid =\lim_{t \rightarrow \pm \infty} \mid I (t)\mid= \frac{\pi}{2}
 \,.\label{eq:properties}
\end{equation}

It can be readily seen that if
\begin{equation}
F>\left(2\gamma \sqrt{\frac{2}{\pi^3}} + \frac{\kappa}{\pi^2} I(t)\right)\cosh\left(\sqrt{\frac{\pi}{2}}\,\frac{\Omega}{2} \right),\label{eq:conditionM}
\end{equation}
then there are roots $\tau_1,\tau_2\in \mathbb{R}$ to the transcendental equations (\ref{eq:Mpp1})-(\ref{eq:Mpm2}). Therefore, the Melnikov functions $M^{(\pm,\pm)}$ and $M^{(\pm,\mp)}$ have zeros. Furthermore, with
\begin{eqnarray}
\frac{\partial M^{(\pm,\pm)}_1}{\partial \tau_1}&=&\mp \tilde{F}\,\cos(\Omega \tau_1)+\tilde{I}\,,\qquad\frac{\partial M^{(\pm,\pm)}_1}{\partial \tau_2}=-\tilde{I}\\
\frac{\partial M^{(\pm,\pm)}_2}{\partial \tau_1}&=&-\tilde{I}\,,\qquad \qquad \qquad\qquad
\frac{\partial M^{(\pm,\pm)}_2}{\partial \tau_2}=\pm\tilde{F}\,
\cos(\Omega \tau_2)-\tilde{I}\,,
\end{eqnarray}
one obtains
\begin{equation}
\det DM^{(\pm,\pm)}(\tau_1,\tau_2)=\pm [\cos(\Omega \tau_1)-\cos(\Omega \tau_2)] \tilde{F} \tilde{I}+\tilde{F}^2 \cos(\Omega \tau_1) \cos(\Omega \tau_2)- \tilde{I}^2\,,
\end{equation}
where we used the abbreviations
\begin{equation}
\tilde{F}=\Omega \sech\left(\sqrt{\frac{\pi}{2}}\,\frac{\Omega}{2} \right),\,\,\,\,{\rm and}\,\,\,\tilde{I}=\frac{\kappa}{\pi^2}\frac{d I(t)}{d t}\,.
\end{equation}
Providing that $\det DM^{(\pm,\pm)}(\tau)\ne 0$ ($\det DM^{(\pm,\mp)}(\tau)\ne 0$) then the homoclinic Melnikov vector $M^{(\pm,\pm)}$ ($M^{(\pm,\mp)}$) has simple zeros implying transversal intersections of the stable and unstable manifolds \cite{Wiggins}-\cite{Yagasaki1}.

With regard to the coupling strength, $\kappa$, the condition in (\ref{eq:conditionM})  imposes also a constraint on the magnitude of the product $\kappa \cdot I(t)$.
Moreover, 
using the properties in (\ref{eq:properties}) one obtains the following homoclinic bifurcation threshold
\begin{equation}
F_c=\left(2\gamma \sqrt{\frac{2}{\pi^3}} + \frac{\kappa}{2\pi}\right)\cosh\left(\sqrt{\frac{\pi}{2}}\,\frac{\Omega}{2} \right).\label{eq:excondition}
\end{equation}
That is, for $F>F_c$ the existence of roots  to the transcendental equations (\ref{eq:Mpp1})-(\ref{eq:Mpm2}) is guaranteed giving birth to homoclinic orbits being of importance for the  existence of strange attractors. 
Note that this bifurcation condition limits  the maximal value of the coupling strength $\kappa$ but is independent of the value of $\Delta \tau$.

With further regard to the influence of the coupling strength $\kappa$ on the behaviour of the stable and unstable manifolds  the condition for the existence of zeros of the Melnikov functions, as given in (\ref{eq:conditionM}), confines the magnitude of the product $\kappa\cdot I(\Delta \tau)$, viz. relates the value of $\kappa$ and $\Delta \tau$. In fact,  taking into account the properties of $I(\Delta \tau)$ (see  Eq.\,(\ref{eq:properties})) one concludes that the larger the coupling strength $\kappa$ the smaller $\Delta \tau$ has to be in order that the value of $I(\Delta \tau)$ complies with the condition (\ref{eq:conditionM}). That is,  with increasing values of $\kappa$ the points where the stable and unstable manifolds intersect each other transversely on the homoclinic orbits  get closer to each other,  giving evidence for the tendency towards synchronisation between the two particle oscillators.
Eventually for $\tau_1=\tau_2$ the ensuing `freezing of the dimensionality' reduces the system to a single driven and damped pendulum (one-and-a-half degree of freedom system) which  impedes the occurrence of hyperchaos as this is necessarily connected with  expansions in several directions in phase space simultaneously (several positive Lyapunov exponents). Nevertheless, for sufficiently small $\kappa$,  simple zeros of the Melnikov functions with $\tau_1 \ne \tau_2$ exist allowing for the occurrence of hyperchaos.

In the limiting case $\kappa \rightarrow \infty$ in conjunction with $\Delta \tau \rightarrow 0$
the two components of each of the homoclinic Melnikov vector degenerate to the Melnikov functions associated with a single pendulum that is damped and periodically driven.
The corresponding Melnikov function
\begin{equation}
M^{\pm}(\tau)=-\gamma 2\sqrt{\frac{2}{\pi^3}}\pm F\,\sech\left(\sqrt{\frac{\pi}{2}}\,\frac{\Omega}{2} \right) \sin(\Omega \tau)
\end{equation}
has simple zeros provided the inequality
\begin{equation}
F> 2 \sqrt{\frac{2}{\pi^3}}\cosh\left(\sqrt{\frac{\pi}{2}}\,\Omega \right)\,\gamma \label{eq:Fgamma}
\end{equation}
is satisfied. Clearly, this type of homoclinic motion cannot represent the source of hyperchaos.

\section{Periodic motions\,--\,Subharmonic Melnikov analysis}\label{section:subharmonic}

With view to coherent particle transport mediated by periodic running solutions we investigate the existence of periodic orbits associated with collective rotational motion of the particles in the presence of the coupling between them where again the perturbative approach is used (see Eqs.\,(\ref{eq:q1eps}),(\ref{eq:q2eps})). For $\epsilon =0$ each unperturbed subsystem supports a one-parameter family of periodic orbits corresponding to rotational motion
\begin{eqnarray}
\left(p_{{k}}^{\pm}(t), q_{{k}}^{\pm}(t)\right)=
\left( \pm \sqrt{\frac{2}{\pi}}\frac{1}{{k}}\dn\left(\frac{\sqrt{2\pi}}{{k}}\, t, {k} \right),
\pm \frac{1}{\pi}\sin^{-1}\left[ \sn \left(\frac{\sqrt{2 \pi}}{{k}} \,t, {k}\right)\right]\right)\,,\label{eq:Jacobi}
\end{eqnarray}
where $\dn$ and $\sn$ are the Jacobi elliptic functions with the elliptic modulus ${k}\in(0,1)$ and with period
\begin{equation}
T_{{k}}=\sqrt{\frac{2} {\pi}}k\,{\K}({k})\,,
\end{equation}
where $\K$ is the complete elliptic integral of the first kind.

We investigate periodic motions near the unperturbed resonant tori ${\cal{T}}_{{k}_1,{k}_2}$, where
\begin{eqnarray}
{\cal{T}}_{{k}_1,{k}_2}^{\pm}&=&\left\{(p_{{k}_1}^{\pm}(t+\tau_1), q_{{k}_1}^{\pm}(t+\tau_1), p_{{k}_2}^{\pm}(t+\tau_2), q_{{k}_2}^{\pm}(t+\tau_2)),\tau_i \in[0,T_{{k}_i}),\,i=1,2, \right.\nonumber\\
 & & \left. \theta= \Omega t\mod(2 \pi),\, t \in {\mathbb{R}}\right\}\,.
\end{eqnarray}

The subharmonic Melnikov functions $M^{m/n}_{1,2}(\tau_1,\tau_2)$ for ${\cal{T}}_{{k}_1,{k}_2}$ are determined by
\begin{eqnarray}
M_i^{m/n}(\tau_1,\tau_2)&=&\int_0^{m_0 T}p_{{k}_i}^{+}(t-\tau_i)\left[-\gamma
p^{+}_{{k}_i}(t-\tau_i)-F\sin(\Omega t)\right.\nonumber\\
&+&\left.\kappa \left( q^{+}_{{k}_{i+1}}(t-\tau_{i+1})-q^{+}_{{k}_i}(t-\tau_{i})\right)\right]dt
\,,\,\,\, i=1,2\,,\,\,\,\tau_{3}=\tau_1\,,\,\,\,{\rm and} \,\,\,\,k_3=k_1.
\end{eqnarray}
Substituting the expression in (\ref{eq:Jacobi}) yields
\begin{eqnarray}
M_1^{m/n}(\tau_1,\tau_2)&=&-\frac{2}{\pi}\frac{\gamma}{{k}_i^2}\int_0^{m_0 T} \dn^2 \left( \frac{\sqrt{2\pi}}{{k}_i}\,(t-\tau_i), {k}_i \right)dt\nonumber\\
&-&\sqrt{\frac{2}{\pi}}\frac{F}{{k}_i}\int_0^{m_0 T}\dn\left( \frac{\sqrt{2\pi}}{{k}}\,(t-\tau_i), {k}_i \right)
\sin(\Omega t)\,dt\nonumber\\
&+&\frac{\sqrt{2}}{\pi}\frac{1}{\pi}\frac{\kappa}{{k}_i}\int_0^{m_0 T} \dn \left( \frac{\sqrt{2\pi}}{{k}_i}\,(t-\tau_i), {k}_i \right)\nonumber\\
&\times&\left\{\sin^{-1}\left[ \sn \left(\frac{\sqrt{2 \pi}}{{k}_{i+1}} \,(t-\tau_{i+1}), {k}_{i+1} \right)\right]
-\sin^{-1}\left[ \sn \left(\frac{\sqrt{2 \pi}}{{k}_{i}} \,(t-\tau_{i}), {k}_{i} \right)\right]
\right\}dt\,,\label{eq:Msub}
\end{eqnarray}
where ${k}_i$ satisfies the resonance condition
\begin{equation}
n_i {\sqrt{\frac{2}{\pi}}}k_i\,\K({k}_i)=m_i T_0=m_i\frac{2\pi}{\Omega}\,,\,\,\,i=1,2\,
\end{equation}
and $m_0$ is the least common multiple of $m_1$ and $m_2$. Using the Fourier series expressions for the Jacobian elliptic functions \cite{Abramowitz}  to evaluate the second integral in Eq.\,(\ref{eq:Msub}) one obtains
\begin{eqnarray}
M_1^{m/n}(\tau_1,\tau_2)&=&-\gamma \frac{4}{\sqrt{2} \pi^{3/2} {k}_1 }\frac{m_0}{m_1}n_1 \E({k}_1)\nonumber\\
&-&F I_1({k}_1,m_0,m_1,n_1=1)\sin(\Omega \tau_1)+\kappa \frac{\sqrt{2}}{\pi^{3/2}{k}_1} I_2(\tau_1,\tau_2)\,,\label{eq:Msub1}
\end{eqnarray}
where $\E({k})$ is the complete elliptic integral of the second type.
In (\ref{eq:Msub1}) the function $I_1({k}_1,m_0,m_1,n_1)$ is non-zero only for $n_1=1$ when it is given by
\begin{equation}
I_1({k}_1,m_0,m_1,n_1=1)=\frac{\sqrt{2\pi } m_0}{\Omega {k}_1 \K({k}_1^{\prime})}\sech \left( \pi m_1 \frac{\K({k}_1)}{\K({k}_1^{\prime})}
\right)\,,
\end{equation}
meaning that no periodic oscillations are 
triggered by the external time-periodic modulation unless $n_1=1$. $\K(k^{\prime})$ is the complete elliptic integral of the first kind with ${k}^{\prime}=\sqrt{1-k^2}$. Furthermore,
in (\ref{eq:Msub1}) the expression $I_2(\tau_1,\tau_2)$ is determined by the integral
\begin{equation}
I_2(\Delta \tau)=
\int_0^{m_0 T} \dn \left( \frac{\sqrt{2\pi}}{{k}_1}\,t, {k}_1 \right) \sin^{-1}\left[ \sn \left(\frac{\sqrt{2 \pi}}{{k}_{2}} \,(t+\Delta \tau), {k}_{2} \right)\right]dt\,.\label{eq:I2}
\end{equation}
Denoting $t=\Delta \tau$, the function $I_2(t)$, has the following properties,
\begin{equation}
I_2(0)=0\,,\,\,\,
I_2(t)=-I_2(-t)\,,\,\,\,
\frac{dI_2(t)}{dt}>0\,,
\,\,\,I_{2,max}(t)=\max_{t \in {\mathbb{R}}}I_2 (t) >0\,,\,\,\,{\rm and} \,\,\,I_{2,min}(t)=\min_{t \in {\mathbb{R}}}I_2 (t) <0
 \,.\label{eq:propertiessh}
\end{equation}

Due to symmetry the second component of the Melnikov vector is determined by
\begin{equation}
M_2^{m/n}(\tau_1,\tau_2;{k}_1,{k}_2,m_0,m_2,n_2)=
M_1^{m/n}(\tau_2,\tau_1;{k}_2,{k}_1,m_0,m_2,n_2)\,.\label{eq:Msub2}
\end{equation}
Furthermore, the Melnikov functions possess the following periodicity
\begin{equation}
M_i^{m/n}(\tau_1,\tau_2)=M_i^{m/n}(\tau_1+l_1 T_{{k}_1},\tau_2+l_2T_{{k}_2})\,,
\end{equation}
with integers $l_1,l_2$.

In particular, if ${k}_1={k}_2={k}$, $m_1=m_2$, and $n_1=n_2$ it holds that $\tau_1-\tau_2=lT_{{k}}$ leading to a vanishing
integral $I_2=0$ in Eq.\,(\ref{eq:I2}). Hence, for collective particle motion being frequency locked to the external time-periodic modulation, and when, apart from a phase difference, the dynamics of the two particles is identical, there is zero average energy exchange between the coupled particles.

To obtain a criterion for the existence of periodic orbits near the unperturbed resonant tori ${\cal{T}}_{{k}_1,{k}_2}$ we note that if
\begin{equation}
F>\left(\gamma \frac{4}{\sqrt{2} \pi^{3/2} {k}_1 }\frac{m_0}{m_1}n_1 \E({k}_1) +
\kappa \frac{\sqrt{2}}{\pi^{3/2}{k}_1} I_{2,max}(t)\right)
\frac{\Omega {k}_1 \K({k}_1^{\prime})}{\sqrt{2\pi } m_0}
\cosech \left( \pi m_1 \frac{\K({k}_1)}{\K({k}_1^{\prime})}
\right)\,,
\end{equation}
then there exist roots $\tau_1,\tau_2\in [0,2\pi/\Omega)$
to the transcendental equations (\ref{eq:Msub1}) and (\ref{eq:Msub2}). Hence, $M^{m/n}=(M_1^{m/n},M_2^{m/n})$ has zeros $(\tau_1,\tau_2)$. Furthermore, as $DM^{m/n}(\tau_1,\tau_2) \ne 0$, the system (\ref{eq:q1}),(\ref{eq:q2}) possesses subharmonic orbits of order $m_0$ near each of the unperturbed resonant tori ${\cal{T}}_{{k}_1,{k}_2}$
\cite{Wiggins}-\cite{Yagasaki1}.

\section{Features of transport}\label{section:features}

In the previous section we proved for weak coupling, weak forcing, and weak damping the existence of periodic motion forced by the external time-periodic modulation facilitating the subharmonic Melnikov method. In particular, it turns out that periodic orbits are characterised by $T=j\,T_0$, with integer $j$, and  $T$ is the period of rotational motion of the particles and $T_0=2\pi/\Omega$ is the period of the external time-dependent modulation. 
Departing from such a perturbational approach  we now investigate the features of running solutions accomplishing coherent particle transport.
Periodic running solutions are characterized by
\begin{equation}
q_i(t+T)=q_i(t)+m_i\,,\,\,\, \dot{q}_i(t+T)=\dot{q}_i(t)\,,\,\,\,i=1,2\,,\label{eq:running}
\end{equation}
with some period $T$ and non-vanishing average velocity
\begin{equation}
 \langle \dot{q}_i\rangle =\frac{1}{T}\int_{0}^{T}dt \dot{q}_i(t)=C_i\ne 0\,,\label{eq:average}
\end{equation}
where a constant $C_i$ is positive (negative) when particle $i$ runs to the right (left), that is $m_i>0$ ($m_i<0$) and $i=1,2$.

As noted in section \ref{section:Melnikov} for in-phase motion $q_1(t)=q_2(t)$, that is for identical dynamics of the two particles, the coupling between the particles vanishes and the dynamics is described by the system of a single particle. Crucially, for our choice of the values of the parameters the single-particle system possesses a strange attractor.
Therefore, in-phase running solutions supported by stable attractors associated with periodic transporting motion are excluded.

As the character of other possible running solutions is concerned we state the following

\vspace*{0.25cm}
\noindent {\bf Theorem:} For the coupled system given in Eqs.\,(\ref{eq:q1}),(\ref{eq:q2}) with $\kappa >0$ and $\gamma>0$, for the periodic running solutions determined by Eqs.\,(\ref{eq:running}) and (\ref{eq:average}) it holds that

\vspace*{0.2cm}
\noindent (i) During one period, $T$, the two particles run over an equal distance, that is $m_1=m_2$\,.

\vspace*{0.2cm}
\noindent (ii) No (nontrivial) periodic motion is possible in the
absence of the time-periodic external modulations, i.e. if $F(t)=0$.

\vspace*{0.2cm}
\noindent (iii) For solutions, being frequency-locked to the external time-periodic modulation, $F(t)=F\sin(\Omega t+\theta_0)$, with period $T_0=2\pi/\Omega$, the distance between the particles performs periodic oscillations due to
\begin{equation}
 q_1(t+T)-q_2(t+T)=q_1(t)-q_2(t)\,,\label{eq:periodicdistance}
\end{equation}
and the period $T$
is determined by
\begin{equation}
T=2l\,T_0\,,
\end{equation}
with integer $l\ge 1$.

\vspace*{0.2cm}
\noindent (iv) The coordinates obey
\begin{equation}
q_1\left(t+\frac{1}{2}T\right)=q_2(t)+k\,,\,\,\,q_2\left(t+\frac{1}{2}T\right)=q_1(t)+k\,,\label{eq:distance1}
\end{equation}
with some integer $k\ne 0$,
and, hence
\begin{equation}
q_i(t+T)=q_i(t)+2k\,,\,\,\,i=1,2\,.\label{eq:distance2}
\end{equation}

\vspace*{0.25cm}

\noindent {\bf Proof:}  Suppose that there exists a period $T$ such that the system (\ref{eq:q1}),(\ref{eq:q2}) has a running solution of the form given in (\ref{eq:running}). Multiplying Eq.\,(\ref{eq:q1}) by $\dot{q}_1$ and Eq.\,(\ref{eq:q2}) by $\dot{q}_2$ and adding the two resulting equations we obtain
\begin{eqnarray}
& & \frac{d}{dt}\left[\frac{1}{2}\dot{q}_1^2+\frac{1}{2}\dot{q}_2^2-\frac{1}{2\pi}\cos(2\pi q_1)-\frac{1}{2\pi}\cos(2\pi q_1)+\frac{\kappa}{2}(q_1-q_2)^2\right]\nonumber\\
&=&-F(t)(\dot{q}_1+\dot{q}_2)-\gamma (\dot{q}_1^2+\dot{q}_2^2)\,,
\end{eqnarray}
where $F(t)=F(t+T_0)=F\sin(\Omega t+\theta_0)$ and $T_0=2\pi/\Omega$.
Integrating over one period, $T$, yields
\begin{eqnarray}
& & \int_{(n-1)T}^{nT} d\left[ \frac{1}{2}\dot{q}_1^2+\frac{1}{2}\dot{q}_2^2-\frac{1}{2\pi}\cos(2\pi q_1)-\frac{1}{2\pi}\cos(2\pi q_1)+\frac{\kappa}{2}(q_1-q_2)^2\right]\nonumber\\
& =&-\int_{(n-1)T}^{nT} dt\left[ F(t)(\dot{q}_1+\dot{q}_2)+\gamma (\dot{q}_1^2+\dot{q}_2^2)\right]\,.\label{eq:integral1}
\end{eqnarray}
With the running solution as given in  (\ref{eq:running}) the expression in (\ref{eq:integral1}) reduces to
\begin{eqnarray}
\int_{(n-1)T}^{nT} d \left[\frac{\kappa}{2}(q_1-q_2)^2\right]&=& -\int_{(n-1)T}^{nT} dt\left[ F(t)(\dot{q}_1(t)+\dot{q}_2(t))+\gamma (\dot{q}_1^2(t)+\dot{q}_2^2(t))\right].\label{eq:integral2}
\end{eqnarray}
We denote the r.h.s. of Eq.\,(\ref{eq:integral2}) by
\begin{equation}
I_n=-\int_{(n-1)T}^{nT} dt\left[ F(t)(\dot{q}_1(t)+\dot{q}_2(t))+\gamma (\dot{q}_1^2(t)+\dot{q}_2^2(t))\right].
\end{equation}

Taking into account the property in Eq.\,(\ref{eq:average}), sinusoidal $F(t)$, and discarding the second negative-definite term under the integral it follows that there exists a constant $\tilde{C}>0$ such that the value for $I_n$ can be bounded from above as follows:
\begin{equation}
I_n \le F\,T\,(\mid C_1 \mid +\mid C_2 \mid)=\tilde{C}={\rm const.}\label{eq:upperbound}
\end{equation}

On the other hand, for the l.h.s. of Eq.\,(\ref{eq:integral2}) we derive
\begin{eqnarray}
\Delta_{\kappa,n}&=&\int_{(n-1)T}^{nT}  d \left[\frac{\kappa}{2}(q_1-q_2)^2\right]\nonumber\\
&=&\frac{\kappa}{2} \left[ \left(q_1(n T)-q_2(n T)\right)^2-\left(q_1((n-1)T)-q_2((n-1)T)\right)^2\right]\nonumber\\
&=&\frac{\kappa}{2}\left\{ \left[\left(q_1(0)+n m_1\right)-\left(q_2(0)+n m_2\right)\right]^2\right.\nonumber\\
&-&\left.\left[\left(q_1(0)+(n-1) m_1\right)-\left(q_2(0)+(n-1) m_2\right)\right]^2\right\}\,.
\end{eqnarray}
Using the notation $\Delta q(0)=q_1(0)-q_2(0)$ and $\Delta m=m_1-m_2$ one arrives eventually at
\begin{equation}
\Delta_{\kappa,n}=[2\Delta q(0) + (2n-1)\Delta m]\Delta m\,.\label{eq:Deltam}
\end{equation}
For $\Delta m \ne 0$  the term $\Delta_{\kappa,n}$ grows monotonically with increasing $n$ and therefore, $\Delta_{\kappa,n}$,  cannot be balanced by the corresponding finite counterpart $I_n$  appearing on the r.h.s. in Eq.\,(\ref{eq:integral2}).
$\Delta_{\kappa,n}$ attains a finite value, i.e. becomes $n$-independent only, if $\Delta m=0$, i.e. $m_1=m_2$ which concludes the proof of statement (i) in the above Theorem.

\vspace*{0.25cm}
\noindent For the proof of part (ii) we note that  $\Delta m=0$ in Eq.\,(\ref{eq:Deltam}) implies
\begin{equation}
\Delta_{\kappa,n}=\int_{(n-1)T}^{nT}  d \left[\frac{\kappa}{2}(q_1-q_2)^2\right]=0\,,\label{eq:vanishing}
\end{equation}
so that due to Eq.\,(\ref{eq:integral2}) the following relation is true
\begin{equation}
\int_{(n-1)T}^{nT} dt\left[ F(t)(\dot{q}_1(t)+\dot{q}_2(t))+\gamma (\dot{q}_1^2(t)+\dot{q}_2^2(t))\right]=0\,,
\end{equation}
which in the absence of the external time-dependent modulation, i.e. $F(t)=0$,
reduces to the condition
\begin{equation}
\int_{(n-1)T}^{nT} dt (\dot{q}_1^2(t)+\dot{q}_2^2(t))=0\,.
\end{equation}
This condition can only be satisfied if $\dot{q}_1=\dot{q}_2=0$, $q_1={\rm const.}$, and $q_2={\rm const.}$ which, however, contradicts the assumption of (nontrivial) periodic solutions. Thus, to obtain periodic solutions $F(t)$ cannot be zero which proves part (ii) of the Theorem.

\vspace*{0.25cm}
\noindent In order to prove the remaining statements (iii) and (iv) we introduce the following
variables
\begin{equation}
 x=q_1-q_2\,,\,\,\,\,y=q_1+q_2\,,
\end{equation}
for the difference and sum of the coordinates $q_1$ and $q_2$ respectively.
The corresponding equations of motion expressed in the new variables read as
\begin{eqnarray}
 \ddot{x}&=&-2\kappa x-\gamma \dot{x}-2\sin(\pi x)\cos(\pi y) \equiv f(x,\dot{x},y)\label{eq:difference}\\
\ddot{y}&=&-\gamma \dot{y}-2F(t)-2\cos(\pi x)\sin(\pi y)\equiv g(x,y,\dot{y})\label{eq:sum}\,.
\end{eqnarray}
While only the second equation (\ref{eq:sum}) contains the time-periodic external modulation, the impact of the coupling, related with the parameter $\kappa$, enters only  the first equation (\ref{eq:difference}) where the corresponding term, $-2\kappa x$,  serves as a harmonic restoring force that keeps the motion of the difference variable $x$ bounded. In contrast, the sum variable $y$ can undergo ongoing rotational motion triggered by the external modulation of period $T_0$, viz. $F(t)=F(t+T_0)$. Accordingly, periodic solutions $y(t)$ of Eq.\,(\ref{eq:sum}) are supposed to be frequency-locked to  (multiples of) the external periodic modulation $F(t)$, and using (\ref{eq:running}) and the proven result from (i), i.e. $m_1=m_2=m$, the periodic running solutions of the difference variable $y(t)$ attain the form
\begin{equation}
y(t+l\, T_0)=y(t)+2m\,,\,\,\,\dot{y}(t+lT_0)=\dot{y}(t)\,,\,\,\,\,{\rm and}\,\,\,l \ge 1\,.\label{eq:locked}
\end{equation}

Given further the reflection symmetry of the difference equation (\ref{eq:difference}), $x\leftrightarrow -x$, symmetric (limit cycle) oscillations of $x(t)$ around $x=0$ of some period $T$ with 
\begin{equation}
x(t+T)=x(t)\,,\,\,\,\dot{x}(t+T)=\dot{x}(t)\,,\label{eq:xT}
\end{equation}
are possible for which the following holds 
\begin{equation}
x\left(t+\frac{T}{2}\right)=-x(t)\,,\,\,\,\,\dot{x}\left(t+\frac{T}{2}\right)=-\dot{x}(t).
\label{eq:xTh}
\end{equation}

In order that Eq.\,(\ref{eq:difference}) supports solutions obeying the relation (\ref{eq:xTh}) its r.h.s., $f(x,\dot{x},y)$,   needs to satisfy the reflection symmetry
\begin{equation}
{f}\left(x\left(t+\frac{T}{2}\right),\dot{x}\left(t+\frac{T}{2}\right),y\left(t+\frac{T}{2}\right)\right)=
-f(x(t),\dot{x}(t),y(t)).\label{eq:cond2}
\end{equation}
Taking into account the r.h.s. of Eq.\,(\ref{eq:difference}) this results in the following condition
\begin{equation}
\cos\left[\pi y\left(t+\frac{T}{2}\,\right)\right] =\cos\left[\pi y\left(t)\right)\right].
\label{eq:condition}
\end{equation}

From  Eq.\,(\ref{eq:locked}) we obtain
$y(t)=y(t+l\,T_0)-2m$ which, after substitution in the r.h.s. of Eq.\,(\ref{eq:condition})
gives
\begin{equation}
\cos\left[\pi y\left(t+\frac{T}{2}\,\right)\right] =\cos\left[\pi y(t+l\,T_0)\right]\,,
\end{equation}
from which follows the relation between the frequencies of the periodic motions in  $x(t)$ and $y(t)$:
\begin{equation}
 T=2l\,T_0\,.\label{eq:period}
\end{equation}

Using the relation $\cos[\pi y(t+2l\,T_0)]=\cos[\pi(y(t)+2\cdot 2m)]=\cos[\pi y(t)]$ one readily  verifies the validity of the condition
\begin{equation}
f\left(x(t+T),\dot{x}(t+T),y(t+T)\right)=f(x(t),\dot{x}(t),y(t))
\,,\label{eq:condx1}
\end{equation}
that has to be satisfied further by the r.h.s of Eq.\,(\ref{eq:difference}) in order  to support  periodic solutions  which comply with Eq.\,(\ref{eq:xT}).

\vspace*{0.25cm}
\noindent Furthermore, given the relations in (\ref{eq:xT}),(\ref{eq:xTh}) and (\ref{eq:period}) it holds that
\begin{equation}
x(t+l\,T_0)=(-1)^lx(t)\,,\,\,\,\dot{x}(t+l\,T_0)=(-1)^l\dot{x}(t)\label{eq:xasym}\\
\end{equation}
for integer $l$.
It remains to verify that the expressions  (\ref{eq:locked}) and (\ref{eq:xasym}) for 
periodic solutions $x(t),y(t)$, that are frequency-locked to the external time-periodic modulation $F(t)$, leave the
 r.h.s. of the sum equation (\ref{eq:sum}) invariant. We obtain
\begin{eqnarray}
g(x(t+l\,T_0),y(t+l\,T_0),\dot{y}(t+l\,T_0))&=&-\gamma \dot{y}(t+l\,T_0)-F(t+l\,T_0)-2\cos[\pi x(t+l\,T_0)]\sin[\pi y(t+l\,T_0)]\nonumber\\
&=&-\gamma \dot{y}(t)-F(t)-2\cos[\pi (-1)^lx(t)]\sin[\pi (y(t)+2m)]\nonumber\\
&=&-\gamma \dot{y}(t)-F(t)-2\cos[\pi x(t)]\sin[\pi y(t)]\nonumber\\
&\stackrel{!}{=}& g(x(t),y(t),\dot{y}(t))\,,
\end{eqnarray}
ensuring invariance 
which completes the proof of the statement in (iii).

\vspace*{0.25cm}
\noindent Finally, combining the expressions in (\ref{eq:locked}) and (\ref{eq:xasym}) one infers that in original coordinates the following holds
\begin{eqnarray}
q_1(t+l\,T_0)&=&q_2(t)+m\\
q_2(t+l\,T_0)&=&q_1(t)+m\\
q_i(t+2l\,T_0)&=&q_i(t)+2m\,,\,\,\,i=1,2\,,
\end{eqnarray}
which concludes the proof of part (iv) of the Theorem.
\hspace{8cm}$\square$

\vspace*{0.25cm}
\noindent {\bf Corollary:} No counterpropagating running states exist as
$m_1$ and $m_2$ can only have the same sign implying that the two particles run in the same direction.

\begin{figure}
\includegraphics[scale=0.5]{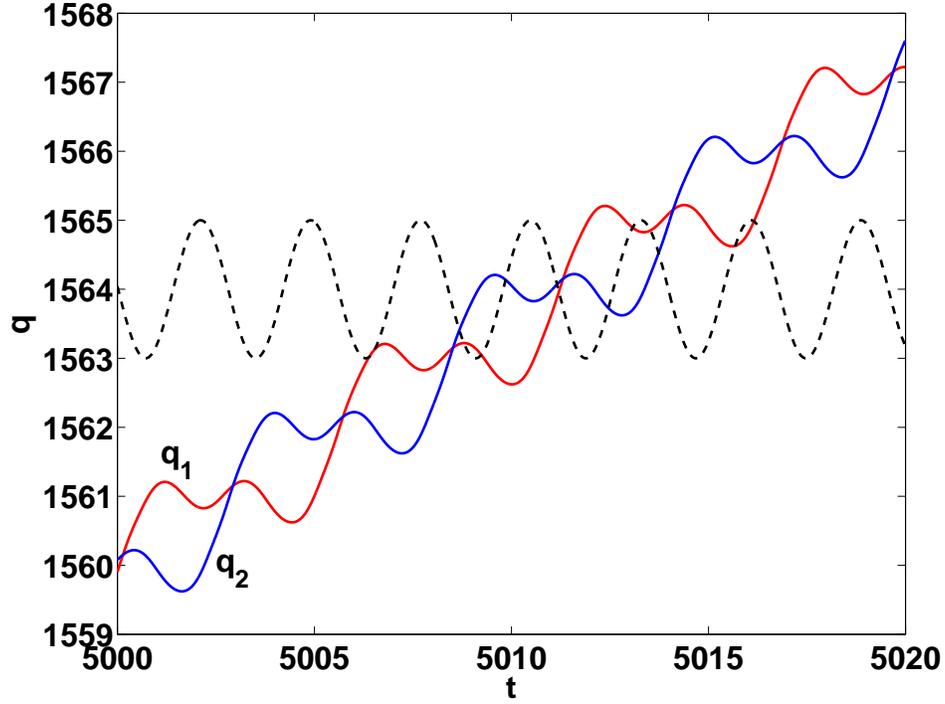}
\caption{Time evolution of the coordinates $q_{1,2}(t)$ of the two particles representing a periodic running state being frequency-locked to the external time-periodic modulation. The parameter values are given by $\Omega=2.25$, $F=1.3$, $\theta_0=0$, $\gamma=0.1$, and $\kappa=0.46$. For comparison, $sin(\Omega t)$ oscillating around $q=1564$ with unit amplitude $F=1$ and frequency $\Omega=2.25$ is shown (dashed line).} \label{fig:running}
\end{figure}

\begin{figure}
\includegraphics[scale=1.5]{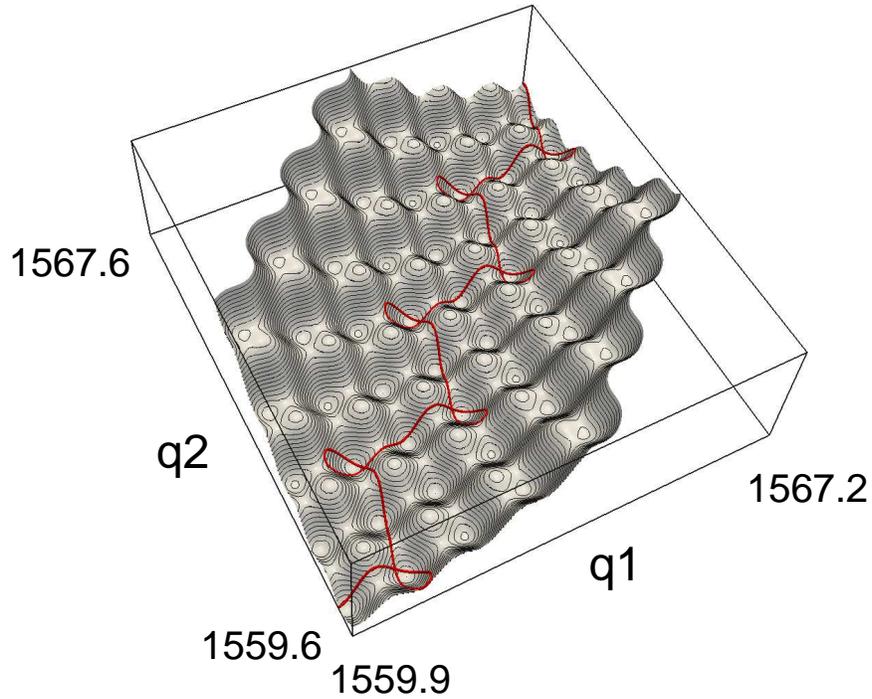}
\caption{Evolution of the trajectory associated with the two particles in a  periodic running states in the spatially periodic potential landscape $U(q_1,q_2)$. The parameter values are the same as in Fig.~\ref{fig:running}.} \label{fig:surf}
\end{figure}

\vspace*{0.25cm} In other words, the theorem states that the periodic transport scenario is characterised by coordinated energy exchange between the two particles proceeding such that they alternatingly surmount the barriers of the periodic potential. We illustrate the features of coherent transport in the periodic regime in Fig.~\ref{fig:running} displaying the time evolution of the coordinates $q_{1,2}(t)$ of the two particles for coupling strength $\kappa=0.46$, that is within the window of periodic motion of the bifurcation diagram (see Fig.~\ref{fig:bif}). To illustrate the frequency-locking between the coordinates and the the time-periodic external modulation, $F(t)=F \sin(\Omega t)$,  the latter, oscillating around $q=1564$ with unit amplitude $F=1$,  is also shown in Fig.~\ref{fig:running}. Fig.~\ref{fig:surf} displays the evolution of the corresponding trajectory in the spatially periodic potential landscape $U(q_1,q_2)=[2-\cos(2\pi q_1)-\cos(2\pi q_2)]/(2\pi)$. Moreover, the temporal behaviour of the coordinates follows the relations given in Eqs.\,(\ref{eq:periodicdistance})-(\ref{eq:distance2}).

In the following we deal with  the existence, uniqueness and approximation of periodic solutions given in (\ref{eq:running}) where $m_1=m_2=m$. To this end we cast the system (\ref{eq:q1}),(\ref{eq:q2}) in matrix notation
\begin{equation}
 \dot{Z}(t)=AZ(t)+B(t,Z(t))\,,\label{eq:matrix}
\end{equation}
where $Z=(p_1,p_2,q_1,q_2)^T$ and the matrix $A$ is given by
\begin{eqnarray}
A =\left( \begin{array}{rrrr}
-\gamma & 0 & -\kappa & \kappa\\
0 & -\gamma & \kappa & -\kappa\\
1 & 0 & 0 & 0\\
0 & 1 & 0 & 0
\end{array} \right)\,.\nonumber
\end{eqnarray}
The inhomogeneity is given by
\begin{equation}
 B(t,Z(t))=(-\sin(2\pi q_1(t))-F(t),-\sin(2\pi q_2(t))-F(t),0,0)^T.\label{eq:BZt}
\end{equation}
The solution to (\ref{eq:matrix}) with initial conditions $Z(0)$ reads as
\begin{equation}
 Z(t)=\exp(At)Z(0)+\int_0^td\tau\exp[A(t-\tau)]B(\tau,Z(\tau))\,.\label{eq:integral3}
\end{equation}
Periodic solutions of period $T$, as given in (\ref{eq:running}) with $m_1=m_2=m$, fulfil
the relation
\begin{equation}
Z(T)=Z(0)+M\,,
\end{equation}
where $M=(0,0,m,m)^T$\,. Substituting $Z(0)=Z(T)-M$ into (\ref{eq:integral3}) yields a nonlinear integral equation for $Z(T)$:
\begin{equation}
Z(T)=(\mathbb{I}-\exp(AT))^{-1}\left[-\exp(AT)M+\int_0^Td\tau\exp[A(T-\tau)]B(\tau,Z(\tau))\right]\,.\label{eq:nonlinear}
\end{equation}
Concerning the solutions $Z(t)$ of the nonlinear integral equation (\ref{eq:nonlinear})  we state the following 

\vspace*{0.2cm} 
\noindent{\bf Theorem:} The integral equation (\ref{eq:nonlinear}) with $T\in[0,t]$ 
possesses a unique solution. Moreover, if a sequence of functions $\{Z_n\}$ is defined inductively by choosing any $Z_0 \in C([0,t];\mathbb{R}^4)$ and setting 
\begin{equation}
Z_{n+1}=(\mathbb{I}-\exp(AT))^{-1}\left[-\exp(AT)M+\int_0^Td\tau\exp[A(T-\tau)]B(\tau,Z_n(\tau))\right]\,,\label{eq:sequence}
\end{equation}
the sequence $\{Z_n\}$ converges uniformly on $I:=[0,t]$ to the unique solution $Z$ of (\ref{eq:nonlinear}).

\vspace*{0.2cm} 
\noindent{\bf Proof:} The proof utilises Banach's
fixed point theorem. We consider the space of continuous functions $Z$ from $I$ to $\mathbb{R}^4$ equipped with an exponentially weighted metric (the Bielecki metric \cite{Bielecki}) 
\begin{equation}
d_{\alpha}(Z,\tilde{Z}):=\sup_{T\in [0,t]}\frac{\parallel Z(T)-\tilde{Z}(T)\parallel}{\exp(\alpha T)}\,,
\end{equation}
where $\parallel \cdot \parallel$ denotes the Euclidean norm on $\mathbb{R}^4$ and $\alpha>0$ is a constant. The corresponding norm is determined by
\begin{equation}
 \parallel Z \parallel_{\alpha}=\sup_{T\in [0,t]}\frac{\parallel Z(T)\parallel}{\exp(\alpha T)}\,.
\end{equation}
$C([0,t];\mathbb{R}^4,\parallel \cdot\parallel_{\alpha})$ is a Banach space.  A solution to (\ref{eq:nonlinear}) is a continuous function $Z: I \mapsto \mathbb{R}^4$.
Let $S$ 
\begin{equation}
 S:\,\,C([0,t];\mathbb{R}^4)\mapsto C([0,t];\mathbb{R}^4)
\end{equation}
be defined by
\begin{equation}
 [SZ](T)=(\mathbb{I}-\exp(AT))^{-1}\left[-\exp(AT)M+\int_0^Td\tau\exp[A(T-\tau)]B(\tau,Z(\tau))\right].
\end{equation}
Then the fixed point of $S$ represents the solution to (\ref{eq:nonlinear}). Hence we must prove that there exist a unique $Z$ such that $SZ=Z$.

Let 
\begin{equation}
 K_1=\sup_{(T,\tau)\in [0,t] \times [0,t]}\parallel \exp(A(T-\tau))\parallel\,,\,\,\, 
K_2=\sup_{T\in [0,t]}\parallel (\mathbb{I}-\exp(A(T)))^{-1}\parallel\,,
\end{equation}
with a suitable matrix norm $\parallel \cdot \parallel$.
Further, using
\begin{eqnarray}
\lvert \sin(2\pi x)-\sin(2\pi y) \rvert \ &=&2\lvert \cos[\pi(x+y)]\sin[\pi (x-y)]\rvert\nonumber\\ 
&\le& 2\lvert \sin[\pi (x-y)] \rvert \le 2\pi \lvert x- y \rvert, 
\end{eqnarray}
implies that
\begin{equation}
\parallel B(T,Z(T))-B(T,\tilde{Z}(T)) \parallel \le 2\pi \parallel Z(T)-\tilde{Z}(T) \parallel\,,\qquad \forall\, T \in [0,t],\,\,\, (Z,\tilde{Z}) \in \mathbb{R}^4,
\end{equation}
and $B(T,Z(T))$ is defined in (\ref{eq:BZt}).
Let $\alpha=2\pi K_1K_2\,\delta$, where $\delta>1$ is an arbitrary constant. 

In order to apply Banach's fixed point theorem we show that $S$ is a contractive map with contraction constant $1/\delta <1$. For any $Z,\tilde{Z} \in C([0,t];\mathbb{R}^4)$ 
one gets
\begin{eqnarray}
d_{\alpha}(SZ,S\tilde{Z})&=&\sup_{T\in [0,t]}\frac{\parallel [SZ](T)-[S\tilde{Z}](T)\parallel}{\exp(\alpha T)}\nonumber\\
&\le&K_1K_2\sup_{T\in [0,t]}\frac{1}{\exp(\alpha T)}\int_0^T d\tau \parallel B(\tau,Z(\tau))-B(\tau,\tilde{Z}(\tau)) \parallel\nonumber\\
&\le& 2\pi K_1 K_2  \sup_{T\in [0,t]}\frac{1}{\exp(\alpha T)} \int_0^T d\tau \parallel Z(\tau)-\tilde{Z}(\tau) \parallel\nonumber\\
&=&2\pi K_1 K_2  \sup_{T\in [0,t]}\frac{1}{\exp(\alpha T)}\int_0^T d\tau \exp(\alpha \tau)\frac{\parallel Z(\tau)-\tilde{Z}(\tau) \parallel}{\exp(\alpha \tau)}\nonumber\\
&\le& 2\pi K_1K_2\,d_{\alpha}(Z,\tilde{Z}) \sup_{T\in [0,t]}\frac{1}{\exp(\alpha T)}\int_0^T d\tau \exp(\alpha \tau)\nonumber\\
&=&\frac{2\pi K_1K_2}{\alpha} d_{\alpha}(Z,\tilde{Z}) \sup_{T\in [0,t]}\left[1-\exp(-\alpha T)\right]\qquad\,\,\,\,\left(\frac{2\pi K_1K_2}{\alpha}=\frac{1}{\delta}\right)\nonumber\\
&\le& \frac{1}{\delta}\left[1-\exp(-\alpha t)\right] d_{\alpha}(Z,\tilde{Z})<\frac{1}{\delta}d_{\alpha}(Z,\tilde{Z})\,.
\end{eqnarray}
Since $\delta >1$ one concludes that $S$ is a contractive map so that Banach's fixed point theorem applies assuring the existence of a unique fixed point $Z$ of $S$. Moreover, from Banach's theorem follows that the sequence $\{Z_n\}$, defined in (\ref{eq:sequence}) converges uniformly in the norm $\parallel \cdot \parallel_{\alpha}$ to that fixed point $Z$.
\hspace{8cm}$\square$

\section{Emergence of a current}\label{section:current}

\begin{figure}
\includegraphics[height=7cm, width=9cm]{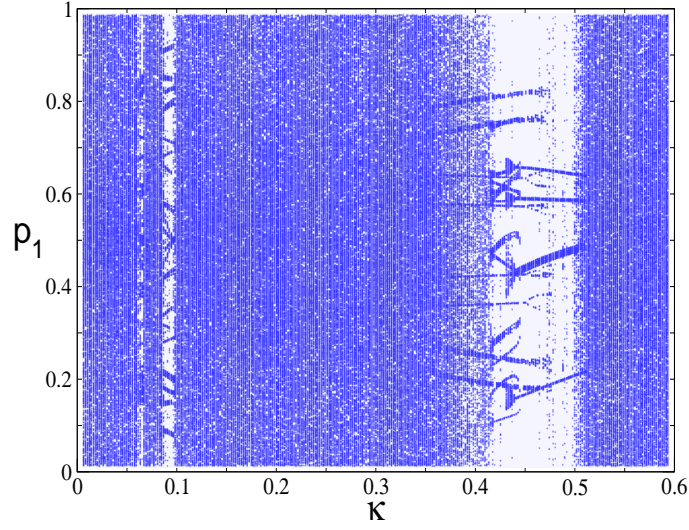}
\caption{\label{fig:bif} (Colour online) Bifurcation diagram as
a function of the coupling strength and remaining parameter values:
$\Omega=2.25$, $F=1.3$, $\theta_0=0$, and $\gamma=0.1$.}
\end{figure}

The dynamics of the coupled particles exhibits very
rich and complex behaviour and depending on the value of the coupling strength $\kappa >0$, one finds periodic, or aperiodic
(quasiperiodic and/or chaotic) solutions in the long time limit.
The character of the phase flow
evolving in a five-dimensional phase space is
conveniently displayed by a Poincar\'{e} map using the period of
the external force, $T_0=2\pi/\Omega$, as the
stroboscopic time. The system of equations of motion was
integrated numerically and omitting a transient phase points were
set in the map at times being multiples of the period duration
$T_0$. In Fig.\,\ref{fig:bif} the bifurcation diagram as a
function of the coupling strength is depicted.

\begin{figure}
\includegraphics[height=8cm, width=10cm]{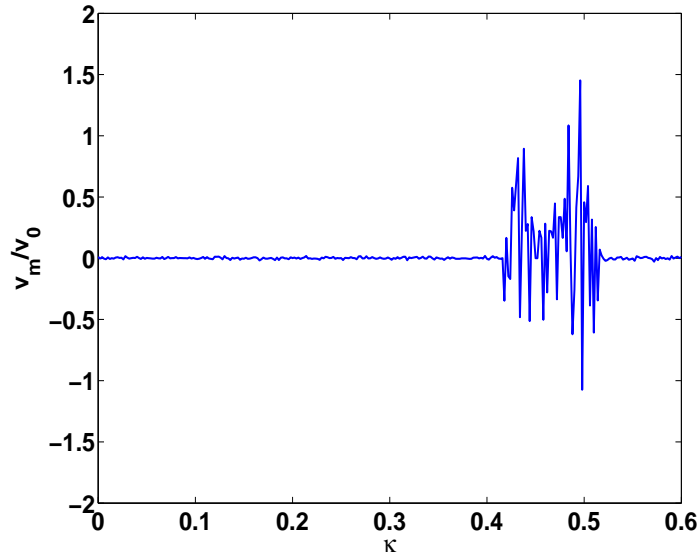}
\caption{\label{fig:current} (Colour online)
Mean velocity $v_m/v_0$ as a function of the coupling strength and the remaining parameter values are as in Fig.~\ref{fig:bif}.}
\end{figure}

Particle transport is quantitatively assessed by the mean velocity, $v_m$, which we
define as the time average of the ensemble averaged velocity, i.e.
\begin{equation}
v_m= \frac{1}{T_s}\,\int_0^{T_s} dt^{\prime} \langle{v}_{1,n}(t^{\prime})
+{v}_{2,n}(t^{\prime})\rangle \,,
\end{equation}
with simulation time $T_s$ and with the ensemble average given by
\begin{equation}
\langle {v}_n(t)\rangle=\frac{1}{N}\sum_{n=1}^N\,{v}_n(t)\,.\label{eq:ensemble}
\end{equation}
Here $N$ denotes the number of particles constituting the ensemble with associated random initial conditions $q_n(0)$ and $p_n(0)$ that are uniformly
distributed over the period of the potential. We express $v_m$ in terms of the ratio of the spatial and temporal
periods $L/T_0 \equiv v_0$ with $v_0\simeq 0.358$ being the
velocity for running solutions that advance by one spatial period
during one period duration of the external field.

We recall that in the uncoupled case, $\kappa=0$, the dynamics is characterised by a strange attractor. Concerning transport, the  associated scenario is that of chaotic running solutions which however,  averaged over time and ensemble,  do not contribute to a non--zero current.

For increasing coupling strength, $\kappa>0$, two typical scenarios arise: pinned and running states.  In the former state the
motion proceeds at most over a finite number of spatial periods
whereas in the latter state motion is directed and unrestricted in
the spatial dimension. In terms of the phase flow, running
asymptotic solutions correspond to phase locked periodic attractors transporting a particle with velocity
$v=m/n$ over $m$ spatial periods of the potential during $n$
period durations $T_0$ of the external periodic field. Running asymptotic solutions may also be supported by aperiodic attractors.

In the bifurcation diagram associated with the dynamics shown in  Fig.\,\ref{fig:bif}
one recognises vertically extended stripes covered densely with
points corresponding to non-phase locked aperiodic attractors and
several periodic windows as well as period-doubling cascades to
chaos. These features of the phase flow are readily attributed to
the resulting mean velocity of the net motion (depicted in
Fig.~\ref{fig:current}).
The ensemble average is taken over an ensemble of $N=5000$
trajectories with uniformly distributed initial conditions $q(0)$ and $\dot{q}(0)=v(0)$. For computation of the long-time average the simulation time
interval for each trajectory is taken as $T_s=5\times 10^5 \simeq 1.8\times 10^5\times
T_0$. We notice almost in the entire $\kappa$-range vanishing mean
velocity $v_m=0$. The exception is the interval  $0.418 \lesssim \kappa \lesssim 0.504$ for which the
solutions are associated with multiple coexisting attractors lying
in  a fairly extended periodic window in Fig.~\ref{fig:bif}. (In the periodic window in the interval $0.082 \lesssim \kappa
\lesssim 0.094$ all solutions are pinned on non-transporting attractors, and thus the  resulting current is zero.)
We observe that the current oscillates wildly with changes of its sign. That is, oppositely running solutions
attributed to branches of the various coexisting periodic attractors contribute to the mean velocity with different weight with the effect that either those with positive, $v=v_0$,  or negative velocity, $v=-v_0$,  dominate  yielding the window of changing mean velocity $v_m$. For increasing $ \kappa \gtrsim 0.504$ these periodic
attractors are destroyed by way of crisis after passage through a
period-doubling route to chaos.

The directed motion results from a
lowering of the dynamical symmetry caused by the external
modulation field \cite{Yevtushenko},\cite{single}. That is, even though the
potential and the external modulation field are with respect to space and time symmetric respectively, with the choice of a  fixed
phase $\theta_0$ the symmetry of the flow is reduced and a
phase-dependent net motion is found. (Note that additional
averaging over the phase $\theta_0$ yields zero mean velocity.)
Due to symmetry reasons it holds that the sign of the mean
velocity is reversed upon the changes $\theta_0=0\,\rightarrow
\theta_0=\pi$ and $F\rightarrow -F$ respectively. However, there
exists a phase $0<\theta_0<\pi$ for which symmetry between the two
coexisting periodic attractors supporting solutions with
velocities of opposite sign, $v_0$ and $-v_0$, is restored and
therefore the net motion vanishes.

\section{Complete synchronisation}\label{section:complete}

In this section we consider the synchronisation features of the coupled system which are related with the properties of the difference system given in Eq.\,(\ref{eq:difference}).

Using the notation $X=(x,\dot{x})^T$ and $F=(0,-2\cos[\pi y]\sin[\pi x])^T$ the solution to Eq.\,(\ref{eq:difference}) with initial conditions $X(0)$ is expressed in form of an integral equation 
\begin{equation}
X(t)=\exp(\Lambda t)X(0)+\int_0^td\tau\exp[\Lambda (t-\tau)]F(\tau,X(\tau))\,,\label{eq:Integral}
\end{equation}
where $\cos[\pi y(t)]$, contained in the inhomogeneity $F$, acts as a time-dependent driving term. The 
matrix $\Lambda$ reads as
\begin{eqnarray}
\Lambda =\left( \begin{array}{rrrr}
0 & 1\\
-2\kappa & -\gamma
\end{array} \right)\,.\label{eq:matrixL}
\end{eqnarray}
The eigenvalues of $\Lambda$ are given by 
\begin{equation}
\lambda_{\pm}=-\frac{1}{2}\left(\gamma \mp \sqrt{\gamma^2-8\kappa}\right)\,,
\end{equation} 
and   fulfil the inequality ${\rm Re}(\lambda_{\pm})<-\Gamma$, $\Gamma>0$ with $\Gamma=(\gamma+\sqrt{\gamma^2-8\kappa}\,)/2$ if $\kappa < \gamma/8$ and $\Gamma=\gamma/2$ if $\kappa \ge \gamma/8$. Hence, there exist a $K>0$ such  that the relation 
\begin{equation}
 \parallel \exp(\Lambda t) \parallel \le K \cdot \exp(-\Gamma t) \,\,\,(t\ge 0)
\end{equation}
holds. $\parallel \cdot \parallel$ is a suitable matrix norm. Further, note that due to 
\begin{equation}
\lvert \cos[\pi y]\sin[\pi x]\rvert \le \lvert \sin[\pi x]\rvert \le \pi  \lvert  x \rvert\,,
\end{equation}
the inhomogeneity $F$ satisfies 
\begin{equation}
\parallel F(t,X(t)) \parallel \le 2 \pi \parallel X(t) \parallel,
\end{equation}
with $\parallel \cdot \parallel$ denoting the Euclidean norm on $\mathbb{R}^2$. 

First, we prove the following theorem.

\vspace*{0.2cm} {\bf Theorem:} Consider the integral equation (\ref{eq:Integral}). If  
\begin{equation}
\Gamma >2\pi K\,,\label{eq:GK}
\end{equation}
then for all initial conditions $X(0)$ the solutions $X(t)$ to (\ref{eq:Integral}) approach zero as $t\rightarrow \infty$.

\vspace*{0.2cm} {\bf Proof:} Concerning the norm of $X$ one gets the following bound
\begin{eqnarray}
\parallel X(t) \parallel &\le& K \exp(-\Gamma t)\parallel X(0) \parallel + K\int_0^t d\tau \exp(-\Gamma(t-\tau)) \parallel F(\tau,X(\tau)) \parallel\nonumber\\
&\le & K \exp(-\Gamma t)\parallel X(0) \parallel + 2\pi K\int_0^t d\tau \exp(-\Gamma(t-\tau)) \parallel X(\tau) \parallel
\end{eqnarray}
Setting $G(t)=\exp(\Gamma t)\parallel X(t) \parallel$ leads to
\begin{equation}
G(t) \le K \parallel X(0) \parallel + 2\pi K\int_0^t d\tau  \, G(\tau) .
\end{equation}
Using Gronwall's inequality we obtain
\begin{equation}
 G(t) \le K \parallel X(0) \parallel \exp\left[2\pi K \int_0^t d\tau\right]=
K \parallel X(0) \parallel \exp\left[2\pi K t\right].
\end{equation}
Therefore,
\begin{equation}
\parallel X(t) \parallel \le K \parallel X(0) \parallel \exp\left[-\left(\Gamma-2\pi K\right)t \right]\,, 
\end{equation}
and as $\Gamma >2\pi K$ it follows that 
\begin{equation}
\lim_{t \rightarrow \infty}\parallel X(t) \parallel =0.\label{eq:asymptotic}
\end{equation}
\hspace{16cm}$\square$

\vspace*{0.2cm} {\bf Corollary:} As $X(t)=(x(t),\dot{x}(t))=(q_1(t)-q_2(t),\dot{q}_1(t)-\dot{q}_2(t))$ one concludes that under the condition (\ref{eq:GK}) for the system of equations (\ref{eq:q1}),(\ref{eq:q2}) the
motion completely synchronises asymptotically and exponentially fast  
regardless of the initial conditions.

\vspace*{0.2cm}
We remark that nothing can be said about the character of the completely synchronised state of the system which can be regular but also chaotic.

\subsection{Transverse stability of the synchronous attractor}

The exchange symmetry properties of the system (\ref{eq:q1}),(\ref{eq:q2}) implies that the hyperspace $q_1=q_2$ and $\dot{q}_1=\dot{q}_2$ constitutes the synchronisation manifold. The latter is invariant, viz. if at any moment of time
the coordinates and velocities of the two particles coincide, then they will do so for all times.  Within the synchronisation manifold the dynamics is governed by the equation for a single particle.

As established in  section \ref{section:model} for our choice of parameter values the dynamics of a single oscillator is characterised by a strange attractor.

In the following we investigate the stability of the synchronisation manifold. To this end
we utilise the difference system given in Eq.\,(\ref{eq:difference}) in terms of which the
synchronisation manifold is represented by the origin $(x,\dot{x})=(0,0)$ in phase space. Therefore, stabilisation of the origin in the difference system is associated with complete (globally identical) synchronisation of the motion of the two particles.

\begin{figure}
\includegraphics[scale=0.5]{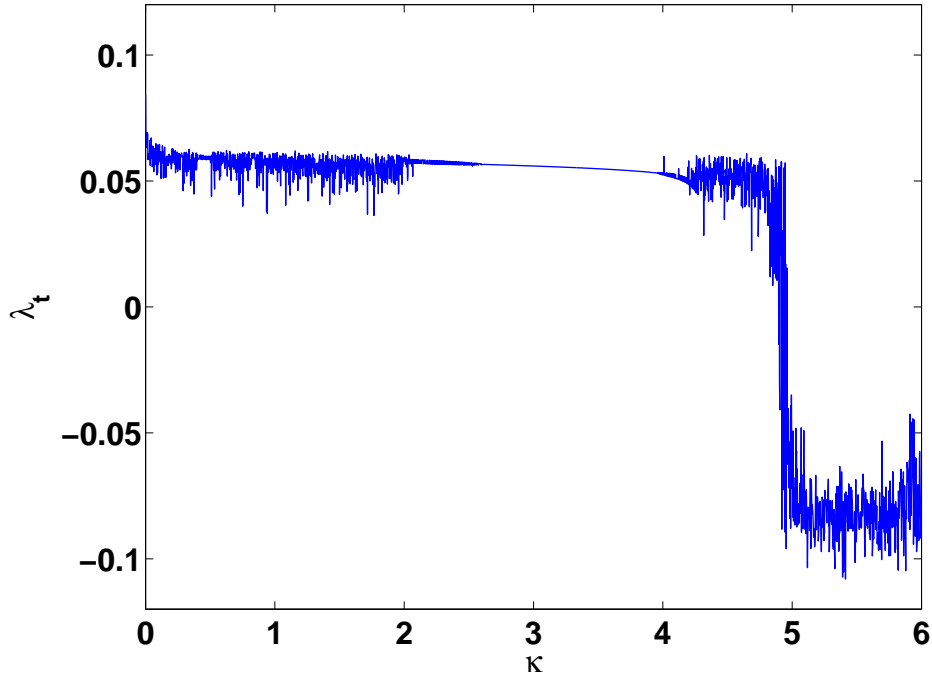}
\caption{The largest transverse Lyapunov exponent in dependence of the coupling strength $\kappa$. Stability of the synchronised chaotic state is assured for $\kappa \gtrsim 5$ for which $\lambda_t$ becomes negative. The remaining parameter values are:
$\Omega=2.25$, $F=1.3$, $\theta_0=0$, and $\gamma=0.1$.} \label{fig:LEt}
\end{figure}

The stability of perturbations transverse to the synchronisation manifold is governed by the largest (transverse) Lyapunov exponent  associated with the difference system  \cite{Fujisaka}. To be precise we study the corresponding tangent equation 
\begin{equation}
\delta \ddot{x}=-2\kappa \delta x-\gamma \delta\dot{x}-2\pi \cos(\pi c(t)) \delta x\,,
\end{equation}
where $c(t)=q_1(t)=q_2(t)$ corresponds to completely synchronous motion. 
In Fig.~\ref{fig:LEt} the largest (transverse) Lyapunov exponent
as a function of the coupling strength is shown.
The largest transverse Lyapunov exponent becomes negative for $\kappa \gtrsim 5$
indicating complete chaotic synchronisation for sufficiently strong coupling strength
in compliance with the effect of `freezing of the dimensionality' occurring for increasing value of $\kappa$ (see Section \ref{section:Melnikov}). Notably, as in the completely synchronised system the coupling between the two particles vanishes, the occurrence of hyperchaos is impossible. Furthermore, although the dynamics in the coherent transport regime, appearing for coupling strengths in the range $0.418 \lesssim \kappa \lesssim 0.504$, is regular, the corresponding transverse Lyapunov exponents are positive as the dynamics does not evolve in the synchronisation manifold and the distance between the particle rather changes periodically according to Eq.\,(\ref{eq:periodicdistance}).

\section{Summary}

We have considered the nonlinear dynamics of two coupled, damped and time-periodically driven  particles evolving in a symmetric and spatially periodic potential.  One aspect of our study  has  concerned the emergence of unstable, irregular motion in the coupled dynamics in the five-dimensional phase space. 
In order to show the existence of transverse homoclinic orbits  we have invoked a homoclinic Melnikov analysis. The latter has enabled  us also to predict the  occurrence of homoclinic bifurcations giving birth to homoclinic orbits being of importance for the  existence of strange attractors. 
Further, we have paid special attention to the influence of the coupling strength on the synchronisation features of the coupled particle system. For relatively weak coupling strengths the dynamics in the five-dimensional phase space may exhibit simultaneous 
expansions in several directions leading to hyperchaos. On the other hand, we have proved that for overcritically strong coupling strengths the motion of the two particles completely synchronises.  The resulting `freezing of the dimensionality' reduces the system to a single driven and damped pendulum, i.e. a one-and-a-half degree of freedom, impeding the emergence of hyperchaos.

Another aspect of our study has dealt with directed particle transport provided by running periodic solutions. Notably, for parameter values for which each uncoupled particle exhibits chaotic dynamics, nevertheless, varying the coupling strength, a fairly extended window
of regular periodic motion appears in the bifurcation diagram. 
As the periodic motion within this window of regular motion is associated with multiple coexisting transporting attractors, 
contributing with different weight to the net flow  a non-vanishing current arises (as a result of associated symmetry-lowering due to the choice of the phase of the external modulation). We have given exact results concerning the features of collective periodic transporting solutions being frequency-locked to the external time-periodic driving. In detail, we have shown that coordinated energy exchange between the particles proceeds  such that they are enabled to 
overcome --\,one particle followed by the other\,-- consecutive barriers of the washboard potential resulting in collective directed motion.

Finally, we remark that it is certainly of interest to extend the present study beyond the case of two coupled particles to infer general features of coherent many-particle transport in periodic landscapes.

\end{document}